\documentclass[12pt]{article}
\usepackage{latexsym}
\usepackage{graphicx}
\usepackage{amsmath}
\usepackage{amsfonts}
\usepackage{amssymb}
\numberwithin{equation}{section}

\newcommand\bea{\begin{eqnarray}}
\newcommand\eea{\end{eqnarray}}
\newcommand\beq{\begin{equation}}
\newcommand\eeq{\end{equation}}

\newcommand{\bib}{\bibitem}
\newcommand{\email}[1]{Electronic mail: \tt #1}
\def\nn{\nonumber}

\def\f{\frac}
\def\la{\langle}
\def\ra{\rangle}

\def\a{\alpha}
\def\d{\delta}
\def\e{\epsilon}
\def\g{\gamma}

\def\G{\Gamma}

\def\om{\omega}

\def\S{\Sigma}

\def\p{\partial} 
\def\CH{{\mathcal{H}}}
\def\CG{{\mathcal{G}}}

\def\n{\eta }
\def\tn{\tilde{\eta} }
\def\tX{\tilde{X}}

\begin{document}
\newcommand{\emaildabhi}{\email{dabhi@rri.res.in}}
\newcommand{\emailroy}{\email{dibyendu@rri.res.in}}
\newcommand{\add}{Raman Research Institute, Bangalore 560080, India.}
\title{Heat transport in harmonic lattices}
\author{Abhishek Dhar\thanks{\add\ \emaildabhi} and Dibyendu
  Roy\thanks{\add\ \emailroy}} 

\maketitle
\begin{abstract}
We work out the non-equilibrium steady state properties of a
harmonic lattice which is connected to heat reservoirs at
different temperatures. The heat reservoirs are themselves modeled as
harmonic systems. Our approach is to write quantum Langevin equations 
for the system and solve these to obtain steady state properties such
as currents and other second moments involving the position and momentum
operators. The resulting expressions will  be seen to be similar in
form to results obtained for electronic transport using the 
non-equilibrium Green's function formalism. As an application of the
formalism we discuss heat conduction in a harmonic chain connected to
self-consistent reservoirs. We obtain a temperature dependent thermal
conductivity which, in the high-temperature classical limit, reproduces the
exact result on this model obtained recently by Bonetto, Lebowitz
and Lukkarinen.    
\end{abstract}
\vskip .5 true cm


\newpage

\section{Introduction}
\label{sec:Intro}
The harmonic crystal is one of the the simplest model that one learns in solid
state physics and it is known to reproduce, for example,  some of the
correct experimental 
features for the specific heat of an  insulating solid. 
The harmonic approximation basically involves expanding the full
atomic potential of the solid about its minimum (which one assumes is
a crystal) and keeping terms up to second order. 
If one transforms to normal mode coordinates then the harmonic crystal
can be viewed as a collection of noninteracting phonons. While many
equilibrium properties can be understood satisfactorily within the
harmonic approximation, transport properties (heat conduction) of the harmonic
lattice are anomalous because of absence of interactions between the
phonons.

In discussions of transport properties (see for example
Ref.~\cite{ziman72}) of an insulating solid it is 
usual to first think of the pure harmonic crystal. One can then
introduce various 
mechanisms for scattering of the phonons, two important ones being  
impurity scattering and  phonon-phonon scattering. 
Impurity scattering can arise because of randomness in the masses of
the particles or in the  spring constants. In this case the phonons
still do not interact with each other. One can think of the 
phonons of the original pure crystal getting elastically scattered by
impurities. Alternatively, since the system is still harmonic, one can
think of heat transmission by the new normal modes of the disordered
system. The second mechanism for scattering is through phonon-phonon
interactions and this occurs if we include the higher order nonlinear
terms ({\emph{i.e}} beyond quadratic order) of the interparticle potential.  
Phonon-phonon interactions are usually classified into those which
conserve momentum and those which do not (Umklapp processes). 

One important question that has attracted a lot of attention is:
 what are the necessary and sufficient conditions for the
validity of Fourier's law of heat conduction 
 \cite{bonetto00,lepri01} ? 
We recall that Fourier's law states that for a solid with a spatially
varying temperature field $T({\bf {x}})$ inside it, the local heat
current density ${\bf{J}}$ at a point ${\bf{x}}$ is given by:
\bea
{\bf{J}}({\bf{x}})=-\kappa \nabla T({\bf{x}})
\eea
where $\kappa$ defines the thermal conductivity of the solid and is
expected to be an intrinsic property of the material. 
Fourier's law is a phenomenological law which is expected to be true
in the hydrodynamic linear response regime. However till now there does
 not exist any purely mechanical model (without external potentials) in which
a first principle demonstration of the validity  of Fourier's law,
 either numerically or analytically, has been achieved. 

Infact it is now pretty much clear that in one dimensional momentum
conserving systems Fourier's law is not valid and one cannot define a
system-size independent thermal conductivity for these systems
\cite{fourier,lepri97,narayan02} ( Note that the momentum one is
referring to here is the
total real momentum, and not the  crystal momentum. This is conserved
if there are no external potentials). Apart from a large number of
numerical studies    
various theoretical approaches have been used for different classes of
systems to arrive at this conclusion regarding non-validity of
Fourier's law in one dimensions.   
In the case of interacting systems (nonlinear) the anomalous behaviour
of thermal conductivity has been understood within the Green-Kubo
formalism and has been related to long-time tails in the current-current
auto-correlation functions \cite{lepri97,narayan02}. For
non-interacting (harmonic) disordered 
systems heat transport occurs through independent phonon
modes and  the main contribution comes from low frequency extended
modes \cite{dhar01}. One finds a situation similar to that of
electronic transport in 
mesoscopic systems \cite{datta,imry} where it is important to include the reservoirs in
discussing transport. In this case it is more natural to talk
of heat conductance of a system and  one finds that this
depends on the details of the reservoirs.

For a three dimensional solid, based on kinetic theory and Boltzmann
equation approaches, the expectation is that the thermal conductivity
should be finite (and Fourier's law valid) at high temperatures where
Umklapp processes dominate. At low temperatures Umklapp processes
are rare and  non-Umklapp ones cannot lead to a finite thermal
conductivity. However it is expected that, if in addition to these
non-Umklapp processes, one also includes impurity scattering , then
one should get a finite conductivity \cite{ziman72}.  Unfortunately
there is no clear proof of any of these physically motivated
expectations. One possible approach to this problem is through a
rigorous study of the phonon Boltzmann equation (see for example
\cite{spohn05}).

In this paper we discuss a formalism for transport in harmonic
lattices based on the Langevin equation approach. This approach was
first used to study heat conduction in a one-dimensional ordered
harmonic lattice \cite{rieder67}. Subsequently this approach was used
to study heat conduction in disordered harmonic lattices in one
\cite{rubin71,connor74,dhar01} and two dimensions \cite{lee05}. 
The quantum mechanical case has also been studied by several authors
\cite{hu87,clel92,chen89,zurcher90,saito00,dhar03,segal03} using an open system
description  either through    quantum Langevin equations or through
density matrices. 
The open system description for harmonic systems closely resembles the
Landauer formalism used for electron transport and
Ref.~\cite{dhar03} gives a  derivation of the basic
Landauer results using the Langevin equation approach, both for
electrons and phonons. Another rigorous approach to studying electron transport
in mesoscopic systems is the non-equilibrium Green's function formalism (NEGF)\cite{datta}
and Ref.~\cite{dhar06} shows how this can be derived, for
non-interacting electrons modeled by tight-binding Hamiltonians, using
a quantum Langevin equation approach.  We note that the Landauer
formalism has also been discussed in the context of wave propagation
in disordered media \cite{sheng,feng}. The problems of heat conduction
in disordered harmonic lattices  and wave propagation in disordered
media are closely related. It is expected that some of the work in
the latter area, for example on localization, will be useful in the
heat conduction context.  

In the present paper we show how  the quantum Langevin equations method
for harmonic lattices leads to NEGF-like expressions for phonon
transport. For simplicity we restrict ourselves to  harmonic
Hamiltonians with scalar displacement variables at each lattice
site. It is straightforward to extend the calculations to the case of vector
displacements.   
The basic steps in the calculation are: (i) one thinks of the full
system as consisting of the sample we are interested in (henceforth
called wire)  as well as  the  reservoirs at different temperatures
which are connected to the wire, (ii) the wire and the reservoir
Hamiltonians are taken to be harmonic, (iii) we eliminate the reservoir
degrees of freedom and this leads to Langevin equations of motion for the wire
variables, (iv) the linear Langevin equations are solved and steady
state properties such as expectation values of the current are
found. Finally (v) the solution is written in a form where one can
identify the usual phonon Green's functions commonly used in solid state
physics. This  leads to the identification with results from the
NEGF formalism. We note that Landauer-like results for phonons
 have been proposed earlier \cite{rego98,blencowe99} and
a recent paper \cite{yamamoto06} derives NEGF results for phonon
transport using the Keldysh approach.

The paper is organized as follows. In Sec.~\ref{sec:QLE} we present
the basic model and derive the quantum Langevin equations describing
the wire.  In Sec.~\ref{sec:Solve} we solve the equations of motion to
obtain the stationary long-time solution which is used in
Sec.~\ref{sec:SSprop} to obtain formal expressions for various steady state
quantities such as  the heat current. In Sec.~\ref{sec:App} we discuss
an application of the present approach to studying heat conduction in
a harmonic wire with self-consistent reservoirs at all lattice
points. Finally we conclude with a discussion in Sec.~\ref{sec:Disc}.

\section{Quantum Langevin equations}
\label{sec:QLE}

We consider a harmonic system which consists of a wire (denoted by
$W$) coupled to reservoirs which are also 
described by harmonic interactions. In most of our discussions we
consider the case of two reservoirs, labeled as $L$ (for left) and
$R$ (right) , which are at two different temperatures. It is easy to
generalize the case where there are more than two reservoirs. 
The Hamiltonian of the entire 
system of wire and reservoirs is taken to be
\bea
{\CH} &=&\f{1}{2}\dot{X}^T M \dot{X}+ \f{1}{2} X^T \Phi X 
\label{fullH} \\
&=&{\CH}_W+{\CH}_L+{\CH}_R+{\mathcal V}_{L}+{\mathcal V}_{R} \nn \\
{\rm where} \quad \CH_W &=&\f{1}{2}\dot{X}_W^T M_W \dot{X}_W+ 
\f{1}{2}X_W^T\Phi_W X_W ~, \nn \\
\CH_L &=&\f{1}{2}\dot{X}_L^T M_L \dot{X_L}+ \f{1}{2}X^T_L \Phi_LX_L ~, \nn \\
\CH_R &=&\f{1}{2}\dot{X}_R^T M_R \dot{X}_R+ \f{1}{2}X_R^T \Phi_R X_R ~, \nn \\
{\mathcal V}_{L}&=&X_W^T V_L X_L,~~~~{\mathcal V}_{R}=X_W^T V_R X_R ~, \nn
\eea
where $M,~M_W,~M_L,~M_R$ are real diagonal matrices representing masses
of the particles in the entire system, wire, left, and right reservoirs
respectively. The quadratic potential energies are given by the real
symmetric matrices $\Phi,~ \Phi_W,~ \Phi_L, ~\Phi_R$ while $V_L$ and $V_R$ denote
the interaction between the wire and the two reservoirs. 
The column vectors $X,~ X_W,~X_L,~X_R$ are Heisenberg operators which
correspond to particle displacements, assumed to be scalars,   
 about some equilibrium configuration. Thus $X=\{ X_1, X_2,...X_{N_s} \}^T$ 
where $X_r$ denotes the position operator of the $r^{\rm{th}}$
particle and $N_s$ denotes the number of points in the entire system.
Also $\dot{X}= M^{-1}~ P$ where $P_r$ denotes the momentum
operator, with $\{X_r,~P_r\}$ satisfying the usual commutation
relations $[X_r, P_s]=i\hbar \delta_{rs}$.

The Heisenberg equations of motion for the system are: 
\bea
M_W \ddot{X}_W = - \Phi_W X_W -V_L X_L -V_R X_R ~,
\eea
and the equations of motion for the two reservoirs are
\bea
M_L \ddot{X}_L &=& -\Phi_L X_L - V_L^T X_W ~, \label{res3} \\
M_R \ddot{X}_R &=& -\Phi_R X_R - V_R^T X_W ~.
\eea
We solve these equations by considering them as linear inhomogeneous
equations. 
Thus for the left reservoir the general solution to Eq.~(\ref{res3})
is (for $t > t_0$):
\bea
X_L(t)&=&f^+_L (t-t_0) M_L X_L(t_0)+ g^+_L (t-t_0) M_L \dot{X}_L
(t_0)\nn \\
&&  - \int_{t_0}^t dt'~ g^+_L (t-t') V_L^T X_W(t')~,~~~~~~\label{solL} \\ 
{\rm with} \quad f^+_L(t)&=&U_L\cos{(\Omega_L t)} U_L^{T}~ \theta(t),~~~
g^+_L(t)=U_L \f{\sin{(\Omega_L t)}}{\Omega_L} U^{T}_L~ \theta(t) ~, \nn
\eea
where $\theta(t)$ is the Heaviside function and $U_L,~\Omega_L $ are the normal mode eigenvector and eigenvalue
matrices respectively and which satisfy the equations:
\bea
U_L^T \Phi_L U_L = \Omega^2_L ~, ~~U_L^T M_L U_L =I~. \nn
\eea
Similarly, for the right reservoir we obtain
\bea
X_R(t)&=&f_R^+(t-t_0) M_R X_R(t_0)+ g_R^+(t-t_0) M_R
\dot{X}_R(t_0)\nn \\&& - \int_{t_0}^t dt'~ g_R^+(t-t') V_R^T X_W(t')~.~~~~ 
\label{solR}
\eea
We plug these solutions back into the equation of motion for the
system to get
\bea
M_W \ddot{X}_W &=& - \Phi_W X_W ~+~ \eta_L ~+~ \int_{t_0}^t dt' ~V_L~ 
g^+_L (t-t')  ~V^T_L~ X_W(t') \nn \\
&& +~ \eta_R ~+~ \int_{t_0}^t dt'~ V_R~ g_R^+(t-t')~  V_R^T~ X_W(t')~,
\label{eqm} 
\eea
 where
\bea
\eta_L &=&-V_L ~[f^+_L (t-t_0)M_L X_L(t_0) ~+~ g_L^+ (t-t_0)
M_L \dot{X}_L (t_0)] \nn \\
\eta_R&=&-V_R~[f_R^+(t-t_0) M_R X_R(t_0) ~+~ g_R^+(t-t_0)M_R \dot{X}_R(t_0)] ~. \label{noiseR}
\eea
This equation has the form of a quantum Langevin equation.
The properties of the noise terms $\eta_L $ and $\eta_R$ are 
determined using the condition that, at time $t_0$, the two isolated
reservoirs are described by equilibrium phonon distribution functions.
At time $t_0$ the left reservoir is in equilibrium at temperature
$T_{L}$ and the population of the normal modes (of the isolated left
reservoir) is given by the
distribution function $f_b(\omega,T_L)=1/[e^{\hbar \omega/k_B T_L}-1]$. The
equilibrium correlations are then given by: 
\bea
\la X_L(t_0) X_L^T(t_0) \ra &=& U_L \f{\hbar}{2 \Omega_L} \coth{(\f{\hbar
\Omega_L}{2 k_B T_L})} U_L^T ~, \nn \\ 
\la \dot{X}_L(t_0) \dot{X}_L^T(t_0) \ra &=& U_L\f{\hbar \Omega_L}{2 } 
\coth{(\f{\hbar \Omega_L}{2 k_B T_L})} U_L^T \nn \\
\la X_L(t_0) \dot{X}_L^T(t_0) \ra &=& U_L (\f{i \hbar}{2}) U_L^T \nn \\
 \la \dot{X}_L^T(t_0) {X_L}(t_0) \ra &=& U_L (\f{-i \hbar}{2}) U_L^T~. \nn 
\eea
Using these  we can determine  the  correlations of the noise terms in
Eq.~(\ref{noiseR}). Thus we get for
the left reservoir noise correlations:
\bea
\la \eta_L(t) \eta_L^T(t') \ra &=& V_L U_L~ \left[\cos{\Omega_L (t-t')} 
\f{\hbar}{2 \Omega_L} \coth{(\f{\hbar \Omega_L}{2 k_B T_L})} \right.\nn \\
 && \left. -i
\sin{\Omega_L (t-t')} \f{\hbar}{2 \Omega_L}\right]~ U_L^{T} V_L^T~,~~~
\label{nono}
\eea
and a similar expression for the right reservoir.

\section{Stationary solution of the equations of motion}
\label{sec:Solve}

Now let us take the limits of infinite reservoir sizes and let $t_0
\to -\infty$. We can then solve Eq.~(\ref{eqm}) by taking  Fourier
transforms. Thus defining the Fourier transforms 
\bea
\tX_W(\omega) &=& \f{1}{2 \pi} \int_{-\infty}^{\infty} dt ~X_W(t) e^{i \omega t} ~,
\nn \\
\tn_{L,R}(\omega)&=& \f{1}{2 \pi} \int_{-\infty}^{\infty} dt~ \n_{L,R}(t) 
e^{i \omega t} ~, \nn \\
g_{L,R}^+(\omega)&=&\int_{-\infty}^{\infty} dt~ g_{L,R}^+(t) e^{i \omega t} ~,
\eea
we get from Eq.~(\ref{eqm})
\bea
(-\omega^2~M_W  +\Phi_W ) \tX_W (\omega) &=& [~\S_L^+ (\omega)+\S_R^+(\omega)~]~ \tX_W(\omega)+\tn_L 
(\omega) + \tn_R (\omega) \nn \\ 
{\rm where} \quad \S_L ^+ (\omega)&=& V_L g_L^+ (\omega) V_L^T ,~~~\S_R^+ 
(\omega)= V_R g_R^+ (\omega)  V_R^T ~. 
\eea
The noise correlations can be obtained from Eq.~(\ref{nono}) and we
get (for the left reservoir):
\bea
\la \tn_L (\omega) \tn_L^T (\omega') \ra &=& \delta (\omega +\omega')~ V_L~ Im [ g_L^+
(\omega) ]~  V_L^T~ \f{\hbar}{\pi} [1+f_b(\omega,T_L)] \nn \\
&=& \delta (\omega +\omega')~ \G_L(\omega)~ \f{\hbar}{ \pi} [1+f_b(\omega,T_L)]~~~\\
&&{\rm where}~~~ \G_L(\omega)=Im[\S_L^+(\omega)] 
\eea
which is a fluctuation-dissipation relation. This also leads to the
more commonly used correlation:
\bea
\f{1}{2}\la~ \tn_L (\omega) \tn_L^T (\omega') +\tn_L (\omega') \tn_L^T (\omega)~ \ra 
= \delta (\omega +\omega') ~\G_L(\omega)~ \f{\hbar}{2 \pi}~{\coth(\f{\hbar \omega}{2k_B
T_L})}.~
\eea
Similar relations hold for the noise from the right reservoir. We
then get the following stationary solution to the equations of motion:
\bea
X_W(t)&=&\int_{-\infty}^\infty d \omega \tX_W (\omega) e^{-i \omega t} ~, \nn \\ 
{\rm with} \quad \tX_W (\omega) &=& G^+_W(\omega)~  {[\tn_L(\omega) + \tn_R (\omega)]}
\label{solSS} ~, \\
{\rm where} \quad G^+_W &=&\f{1}{[-\omega^2 M_W+  
 \Phi_W -\S_L^+ (\omega)-\S_R^+(\omega)]}~.  
\eea
The identification of $G^+_W(\omega)$ as a phonon Green function, with
 $\S^+_{L,R}(\omega)$ as effective self energy terms, is the main step that
enables a comparison of results derived by the quantum Langevin
 approach with those obtained from the NEGF method. In
App.~\ref{app:GF} we show explicitly how this identification is made. 

For the reservoirs we get, from Eqs.~(\ref{solL}-\ref{solR}),
\bea
-V_L \tX_L(\omega) &=& \tn_L (\omega) +\S_L^+ \tX_W(\omega) ~, \nn \\
-V_R \tX_R(\omega) &=& \tn_R(\omega) +\S_R^+ \tX_W(\omega) ~.
\label{solLR}
\eea

\section{Steady state properties} 
\label{sec:SSprop}

{\bf{Current}}: The simplest way to evaluate the steady state current is to 
evaluate the following expectation value for left-to-right current:
\bea
J &=&-\la~ \dot{X}_W^T V_L X_L~ \ra 
=\int_{-\infty}^\infty d \omega \int_{-\infty}^\infty d \omega' ~e^{-i(\omega
+\omega') t} i \omega~ \la \tX_W^T(\omega) V_L \tX_L (\omega') \ra ~, \nn
\eea
which is just the rate at which the left reservoir does work on the
wire. Using the solution in Eq.~(\ref{solSS}-\ref{solLR}) we get 
\bea
J&=& -\int_{-\infty}^\infty d \omega \int_{-\infty}^\infty d \omega'~ e^{-i(\omega
+\omega') t} i \omega ~\la~ (~\tn_L^T(\omega)+\tn_R^T(\omega)~)~
{G^+_W}^T(\omega) \nn \\
&&\times~[~ \tn_L (\omega') 
 + \S_L^+ (\omega')~ G^+_W(\omega')~  (~\tn_L (\omega') + \tn_R(\omega')~)~]~\ra ~.
\eea
Now consider that part of $J$, say $J^{R}$, which depends only on
$T_R$. Clearly this is:
\bea
J^R &=&-\int_{-\infty}^\infty d \omega\int_{-\infty}^\infty d \omega'~ e^{-i(\omega
+\omega') t}~ i~ \omega ~\nn \\ && \times Tr~ [~ {G^+_W}^T(\omega) ~\S_L^+ (\omega')~
  G^+_W(\omega')~  \la~ \tn_R (\omega') \tn_R^T(\omega)~ \ra~ ] \nn \\
=&-&i\int_{-\infty}^\infty d \omega~ Tr[~  {G^+_W}^T(\omega)~ \S_L^+ (-\omega)
G^+_W(-\omega) \G_R (\omega)]~ \f{\hbar \omega}{ \pi}~ 
[1+f_b(\omega,T_R)] ~. \nn
\eea
Using the identities ${G_W^+}^T= G_W^+ $, $G_W^+(-\omega)=G_W^-(\omega)$ and
taking the real part of above equation we obtain, after some simplifications,
\bea
J^R=-\int_{-\infty}^\infty d \omega~ Tr[~ {G^+_W}(\omega)~  \G_L (\omega)~ 
G^-_W(\omega)~ \G_R (\omega)~]~ \f{\hbar \omega}{ \pi}~ [1+f_b(\omega,T_R)] ~.\nn
\eea
Including the contribution from the terms involving $T_L $, and noting
that the current has to vanish for $T_L=T_R$, it is clear
that the net current will be given by
\bea
J= \int_{-\infty}^\infty d \omega~ Tr[~ {G^+_W}(\omega)~  \G_L (\omega) 
G^-_W(\omega) ~ \G_R (\omega)]~ \f{\hbar \omega}{ \pi}~ [f(\omega,T_L)-f(\omega,T_R)] ~.\label{SSJ}
\eea
This expression for current can be seen to be of identical form as
the NEGF expression for electron current (see for example
\cite{datta,meir92,kamenev04}).

{\bf {Two point correlation functions}}: We can also easily compute
expectation values of various correlations.  
Thus the velocity-velocity correlations are given by 
\bea
K&=& \la \dot{X}_W \dot{X}_W^T \ra \nn \\
&=& \int_{-\infty}^\infty d \omega~ \f{\omega}{\pi}~ [~ G^+_W(\omega)  \G_L (\omega)
G^-_W(\omega)  {\hbar \omega} (1+f_b(\omega,T_L)) \nn \\
&&~~~~~~~~~~~~ + G^+_W(\omega)  \G_R (\omega) G^-_W(\omega)  
{\hbar \omega} (1+f_b(\omega, T_R))~] ~ \nn \\
&=& \int_{-\infty}^\infty d \omega \f{\omega}{\pi}~ [~ G^+_W(\omega)  \G_L (\omega)
G^-_W(\omega)  \f{\hbar \omega}{2} \coth(\f{\hbar \omega}{2k_BT_L}) \nn \\
&&~~~~~~~~~~~~ + G^+_W(\omega)  \G_R (\omega) G^-_W(\omega)  
\f{\hbar \omega}{2} \coth(\f{\hbar \omega}{2k_BT_R}) ~]~,
\label{Kneq}
\eea
where the last line is easily obtained after writing $K=(K+K^*)/2$. 
We see that for $T_L=T_R$ this reduces to the equilibrium result of
Eq.~(\ref{eqK}) {\emph {provided that there are no bound states}}.

Similarly the position-position and position-velocity correlations are
given by:
\bea
P&=& \la {X_W} X_W^T \ra \nn \\
 &=& \int_{-\infty}^\infty d \omega \f{\hbar}{2 \pi}~ [~ G^+_W(\omega)  \G_L (\omega)
G^-_W(\omega)   \coth(\f{\hbar \omega}{2k_BT_L}) \nn \\
&&~~~~~~~~~~~~ + G^+_W(\omega)  \G_R (\omega) G^-_W(\omega)  
 \coth(\f{\hbar \omega}{2k_BT_R}) ~]~,\nn \\
C &=&  \la {X_W} \dot{X}_W^T \ra \nn \\
&=& \int_{-\infty}^\infty d \omega \f{i}{\pi}~ [~ G^+_W(\omega)  \G_L (\omega)
G^-_W(\omega)  \f{\hbar \omega}{2} \coth(\f{\hbar \omega}{2k_BT_L}) \nn \\
&&~~~~~~~~~~~~ + G^+_W(\omega)  \G_R (\omega) G^-_W(\omega)  
\f{\hbar \omega}{2} \coth(\f{\hbar \omega}{2k_BT_R}) ~]~. \label{qucorr}
\eea 
The correlation functions $K$ and $P$ can be used to define the local
energy density which can in turn be used to define the 
temperature profile in the non-equilibrium steady state of the
wire. Also we note that the correlations $C$ give the local heat
current density. In the next section we will find that it is sometimes
more convenient to evaluate the total steady state current from this
expression  rather than the one in 
Eq.~(\ref{SSJ}).

{\bf{Classical limits}}: The classical limit is obtained by taking
the high temperature limit 
so that $\hbar \omega/k_B T \to 0$. Then we obtain the following
expressions for the various steady state properties computed in the
last section. The current is given by
\bea
J=  \f{k_B ~(T_L- T_R)}{\pi} ~\int_{-\infty}^\infty d \omega~ Tr[~ {G^+_W}(\omega)~  \G_L (\omega) 
G^-_W(\omega) ~ \G_R (\omega)]~, \label{cle}
\eea
while other correlation functions are given by:
\bea
K
&=& \f{k_B T_L}{\pi} \int_{-\infty}^\infty d \omega ~\omega ~ G^+_W(\omega)  \G_L (\omega)
G^-_W(\omega) \nn \\&& +  \f{k_B T_R}{\pi} \int_{-\infty}^\infty d \omega ~\omega~
G^+_W(\omega)  \G_R (\omega)  G^-_W(\omega)~,\nn \\   
P &=& \f{k_B T_L}{\pi} \int_{-\infty}^\infty d\omega~ \f{1}{\omega} ~ G^+_W(\omega)  \G_L (\omega)
G^-_W(\omega) \nn \\&& +  \f{k_B T_R}{\pi} \int_{-\infty}^\infty d \omega ~\f{1}{\omega}~
G^+_W(\omega)  \G_R (\omega)  G^-_W(\omega)~,\nn \\   
C &=& \f{i k_B T_L}{\pi} \int_{-\infty}^\infty d \omega ~ G^+_W(\omega)  \G_L (\omega)
G^-_W(\omega)  \nn \\&&+  \f{i k_B T_R}{\pi} \int_{-\infty}^\infty d \omega ~
G^+_W(\omega)  \G_R (\omega)  G^-_W(\omega)~.  \label{clcorr} 
\eea 
For one dimensional wires these lead to \cite{dhar03} expressions for
current and 
temperature used in earlier studies of heat conduction in disordered
harmonic chains \cite{rubin71,connor74,dhar01}. 

\section{An application: one-dimensional harmonic crystal with self-consistent
  reservoirs} 
\label{sec:App}
As an application of the Langevin equation-Green functions formalism
we consider the problem of heat transport in a harmonic
chain with each site connected to self-consistent heat
reservoirs. The classical version of this model was first studied by
\cite{viss70,viss75} who introduced the self-consistent
reservoirs as a simple scattering mechanism which might ensure local
equilibration and  the validity of Fourier's law.  
This model was recently solved exactly by Bonetto et al \cite{bonetto04}
who proved local equilibration and validity of Fourier's law and
obtained an expression for the thermal conductivity of the wire. They
also showed that the temperature profile in the wire was linear. 
The quantum version of the problem was also studied by Visscher and
Rich \cite{viss75b} who analyzed the limiting case of weak coupling to the
self-consistent reservoirs.   
We will show here how the present formalism can be used to obtain results
in the quantum-mechanical case. The classical results of Bonetto et al
are obtained as the high temperature limit while the quantum mechanical
results of Vischer and Rich are obtained in the weak coupling limit.

In this model one considers a  one-dimensional harmonic wire
described by the Hamiltonian
\bea
H_W &=& \sum_{l=1}^N \f{m}{2} [ \dot{x}_l^2 +  \omega_0^2 x_l^2] +\sum_{l =1}^{N+1}
\f{m\om_c^2}{2} (x_l-x_{l-1})^2~, \nn \\
   &=& \f{1}{2} \dot{X}_W^T M_W \dot{X}_W +\f{1}{2} X^T_W \Phi_W X_W 
\eea
where the wire particles are denoted as $X^T=\{x_1,x_2,...x_N\}$ and
we have chosen the boundary conditions $x_0=x_{N+1}=0$.  
All the particles are connected to heat reservoirs which are taken to
be Ohmic. The coupling strength to the reservoirs is
controlled by the 
dissipation constant $\g$. The temperatures of the first and last
reservoirs are fixed 
and taken to be $T_1=T_L$ and $T_N=T_R$. For other particles,
{\emph{i.e}}  $l=2,3...(N-1)$, the temperature of the attached
reservoir $T_l$ is fixed self-consistently in such a way that the net
current flowing into any of the reservoirs $l=2,3...(N-1)$ vanishes.  
The Langevin equations of motion for the particles on the wire are:
\bea
m \ddot{x}_l&=&-m \om_c^2 (2 x_l -x_{l-1}-x_{l+1})-m \omega_0^2
x_l-\gamma \dot{x}_l +\n_l~~~~{l=1,2...N}~,
\eea
where the noise-noise correlation is easier to express in frequency
domain and given by
\bea
\f{1}{2}  \la~ \n_l (\om) \n_m(\om')  + \n_l (\om') \n_m(\om)~ \ra 
&=& \f{\g \hbar \omega}{2 \pi} \coth (\f{\hbar
  \omega}{2 k_B T_l})~\d (\om+\om')~\d_{lm} ~.\label{qnnw}
\eea
From the equations of motion it is clear that the $l^{\rm th}$ particle is
connected to a bath with a self energy matrix $\Sigma^+_l(\omega)$  whose only
non vanishing element is $[\Sigma^+_l]_{ll}=i \g \omega$.
Generalizing Eq.~(\ref{SSJ}) to the case of multiple baths we find
that the heat current from the $l^{\rm th}$ reservoir into the wire is
given by: 
\bea
J_l&=& \sum_{m=1}^N ~\int_{-\infty}^\infty d \omega~ Tr[~
  {G^+_W}(\omega)~  \G_l (\omega) G^-_W(\omega) ~ \G_m (\omega)]
\f{\hbar \om}{\pi}~[f(\om,T_l)-f(\om,T_m)]~,~~ \label{JSCR} \\
&&{\rm{where}}~~~G^+_W= [~-\omega^2~M_W +\Phi_W -\sum_l \S^+_l(\omega)~]^{-1}~,~~~\G_l=
Im[\S^+_l]~. \nn 
\eea
Using the form of $\G_l$ we then get:
\bea
J_l= \sum_{m=1}^N  \g^2 ~\int_{-\infty}^\infty d
\omega~ \omega^2 \mid [G^+_W(\omega)]_{lm} \mid^2 ~\f{\hbar
  \omega}{\pi}~[f(\omega,T_l)-f(\omega,T_m)] ~ . \nn
\eea
To find the temperature profile we need to solve the $N-2$ nonlinear
equations $J_l=0$ for $l=2,3...N-1$ with $T_1=T_L$ and $T_N=T_R$. 
To proceed we consider the linear response regime with the applied
temperature difference $\Delta T = T_L-T_R << T$ where
$T=(T_L+T_R)/2$. In that case we expand the phonon distribution
functions $f(\om,T_l)$ 
about the mean temperature $T$ and get the following simpler expressions for
the currents
\bea
J_l= \g^2 ~\int_{-\infty}^\infty d
\omega~ \f{\hbar \omega^3}{\pi} \f{\p f(\omega,T)}{\p T}~
	\sum_{m=1}^N ~\mid [G^+_W(\omega)]_{lm} \mid^2 ~(T_l-T_m)~. 
\eea  
We write $G^+=Z^{-1}/(m\omega_c^2)$ where $Z$ is a tridiagonal matrix
with offdiagonal elements equal to $-1$ and diagonal elements are all
equal to $z=2+\om_0^2/\om_c^2-\om^2/\om_c^2-i \g \om/(m \om_c^2)$.
It is then easy to find its inverse using the formula
$Z^{-1}_{lm}= {D_{1,l-1} ~D_{m+1,N}}/{D_{1,N}}$,   
where $D_{ij}$ is the determinant of the sub-matrix of $Z$ beginning
with the $i$th row and column and ending with the $j$th row and
column. The determinant is given by $D_{ij}=
\sinh{[(j-i+2)~\a]}/\sinh(\a)$, where $e^{\a}= z/2 \pm [(z/2)^2
  -1]^{1/2}$ (any of the two roots can be taken).
For points far from the  boundaries of the wire ($l=y N$ where
$y=O(1),~1-y=O(1)$) we then find that  
\bea
G^+_{lm}= \f{e^{-\a |l-m|} }{2 m \om_c^2 \sinh{\a}} \label{Gasym}~,
\eea
where we choose the root $\a$ such that $\a_R=Re[\a] > 0$. 
In Ref.~\cite{bonetto04}, where the classical version of the present
model was studied, it was shown that in the limit $N\to \infty$
the temperature profile obtained by solving the self-consistent equations
has the linear form 
\bea
T_l=T_L+\f{l-1}{N-1} (T_R-T_L)~.
\eea
From the form of $G^+_{lm}$ in Eq.~(\ref{Gasym}) we see at once that, for
any point $l$  in the bulk of the wire, the zero-current condition
$J_l=0$ is satisfied since  $\sum_{m=-\infty}^\infty (l-m) |e^{-\a |l-m|}|^2 =0$. 
For points which are within distance $O(1)$ from the boundaries the
temperature profile deviates from the linear form. 

To find the net left-right current in the wire we could use the
formula for $J_1=-J_N$ given in Eq.~(\ref{JSCR}). However we notice
that use of this formula requires us to know the accurate form of 
$T_m$ for points $m$ close to the boundaries since these terms 
contribute significantly to the sum in Eq.~(\ref{JSCR}). We instead
use a different expression for the current. We evaluate the left-right
current  $J_{l,l+1}$ on the bond connecting sites $l$ and $(l+1)$. Using
Eq.~(\ref{qucorr}) and in the linear response limit, making
expansions about $T$, we get   
\bea 
J_{l,~l+1}&=& m \om_c^2~\la x_l \dot{x}_{l+1} \ra = -\f{ m \om_c^2 \g
}{\pi}  \int_{-\infty}^\infty d \omega ~
\omega~ 
\left(\f{\hbar \om}{2 k_B T}\right)^2 {\rm{cosech}}^2 (\f{\hbar
  \om}{2k_B T})~ \nn \\
&& \times \sum_{m=-\infty}^{\infty} k_B T_m ~Im \{ [G^+_W(\omega)]_{lm}
     [G^+_W(\omega)]^*_{l+1~m}\}   ~.\nn
\eea
The current value is independent of $l$ and we  choose to evaluate it
at a value of $l$ in the bulk of the wire. In that case terms in the
sum above which contain $T_m$ with $m$ close to the boundaries are
exponentially small ($\sim e^{-\a N}$) and so do not contribute. Hence
we can use the linear temperature profile for $T_m$ and the form of
$G^+_W$ in Eq.~(\ref{Gasym}) to evaluate the heat current. We get:
\bea
J &=&-\f{  \g
}{8 m \om_c^2 \pi i}  \int_{-\infty}^\infty d \omega ~
\f{\omega}{|\sinh{\a}|^2}~ 
\left(\f{\hbar \om}{2 k_B T}\right)^2 {\rm{cosech}}^2 (\f{\hbar
  \om}{2k_B T})~ \nn \\
&& \times \sum_{m=-\infty}^\infty {k_B T_m}~[e^{-\a |l-m|}e^{-\a^*
    |l+1-m|} -   e^{-\a^* |l-m|}e^{-\a^ |l+1-m|}  ]~, \nn
\eea  
where the additional terms in the summation over $m$ are
exponentially small contributions. Also as before we choose the root
$\a (\omega)$ such that $Re[\a] > 0$. Using the notation
$\a_R(\omega)=Re[\a],~\a_I(\omega)=Im[\a]$ we get, after some algebra,
the following expression for the thermal conductivity
$\kappa=JN/\Delta T$ (obtained in the large $N$ limit): 
\bea
\kappa=\f{\g k_B }{16 m \om_c^2 \pi i} \int_{-\infty}^{\infty} d \omega
~\f{\omega}{\sinh^2{\a_R}}\left(\f{\hbar \om}{2 k_B T}\right)^2 {\rm{cosech}}^2 (\f{\hbar
  \om}{2k_B T})  ~ \left( \f{1}{\sinh{\a}}-\f{1}{\sinh{\a^*}} \right)~.\label{genF}
\eea
In the high temperature limit we get
$(\hbar \om/2 k_B T)^2 {\rm{cosech}}^2(\hbar \om/2k_B T)~\to~1$. This
gives the classical result for the thermal conductivity.  In this
limit  a
change of variables from $\omega$ to $\a_I$ leads to the following
result for the thermal conductivity:
\bea
\kappa_{cl} &=&\f{2 k_B m\om_c ^2(2+\nu^2)}{\g \pi} \int_0^{\pi/2} d \a_I~ \f{\sin^2
  {(\a_I)}}{(2+\nu^2)^2 -4 \cos^2{(\a_I)}} \nn \\
&=&\f{k_B m \om_c^2}{\g~(2+\nu^2+[\nu^2(4+\nu^2)]^{1/2})}~,  
\eea
where $\nu=\om_0/\om_c$.
This agrees with the result obtained in Ref.~\cite{bonetto04}.

An interesting limiting case is the case of weak coupling 
to the reservoirs ($\gamma \to
0$). In this case Eq.~(\ref{genF})
gives:
\bea
\kappa_{wc} &=& \left(\f{\hbar \om_c^2}{k_B T}\right)^2 \f{m k_B}{4 \gamma
  \pi} \int_0^{\pi}~ d \a_I~ \sin^2{\a_I}~
      {\rm{cosech}}^2 (\f{\hbar \om_\a}{2 k_B T}) ~,\label{wckappa}\\ 
{\rm where}~~~ \om_\a^2 &=&\om_0^2+2 \om_c^2 [1-\cos(\a_I)] \nn~.
\eea
This agrees with the result obtained in \cite{viss75b}. The
temperature profile obtained by them differs from the linear form
obtained by us and also by \cite{bonetto04} and is 
incorrect. They nevertheless obtain the correct thermal conductivity,
presumably because their derivation only uses the temperature profile close to
the centre of the chain where it is again linear.
In the low
temperature limit, Eq.~(\ref{wckappa}) gives $\kappa_{wc} \sim
e^{-\hbar \om_0/k_B T}/T^{1/2} $ for $\om_0 \neq 0$ and $\kappa_{wc} \sim T$ for 
$\om_0=0$.
As noted in \cite{viss75b} the expression for thermal
conductivity (in the weak scattering limit) is consistent with a
simple relaxation-time form for the thermal conductivity. The
temperature dependence  of $\kappa_{wc}$ then simply follows the
temperature dependence of the specific heat of the $1$-dimensional
chain. 

\begin{figure}[htb]
\vspace{1cm}
\includegraphics[width=12.5cm]{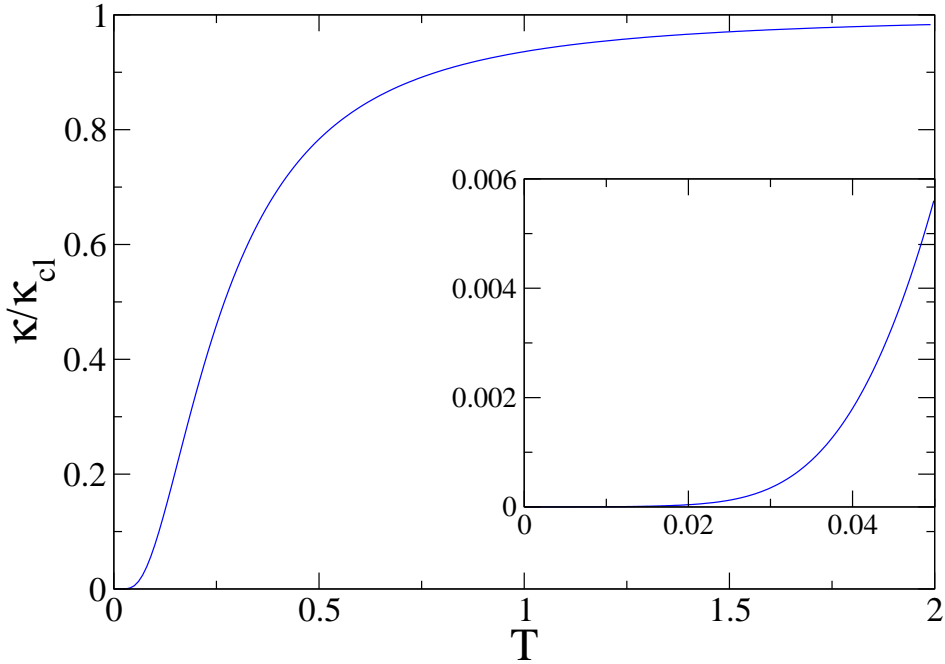}
\caption{ Plot of the scaled thermal conductivity as a function of
  temperature (in units of $\hbar \om_c/k_B$)  for $\nu=0.5$. Inset
  shows the low temperature behaviour. } 
\label{fig1}
\end{figure}

\begin{figure}[htb]
\includegraphics[width=12.5cm]{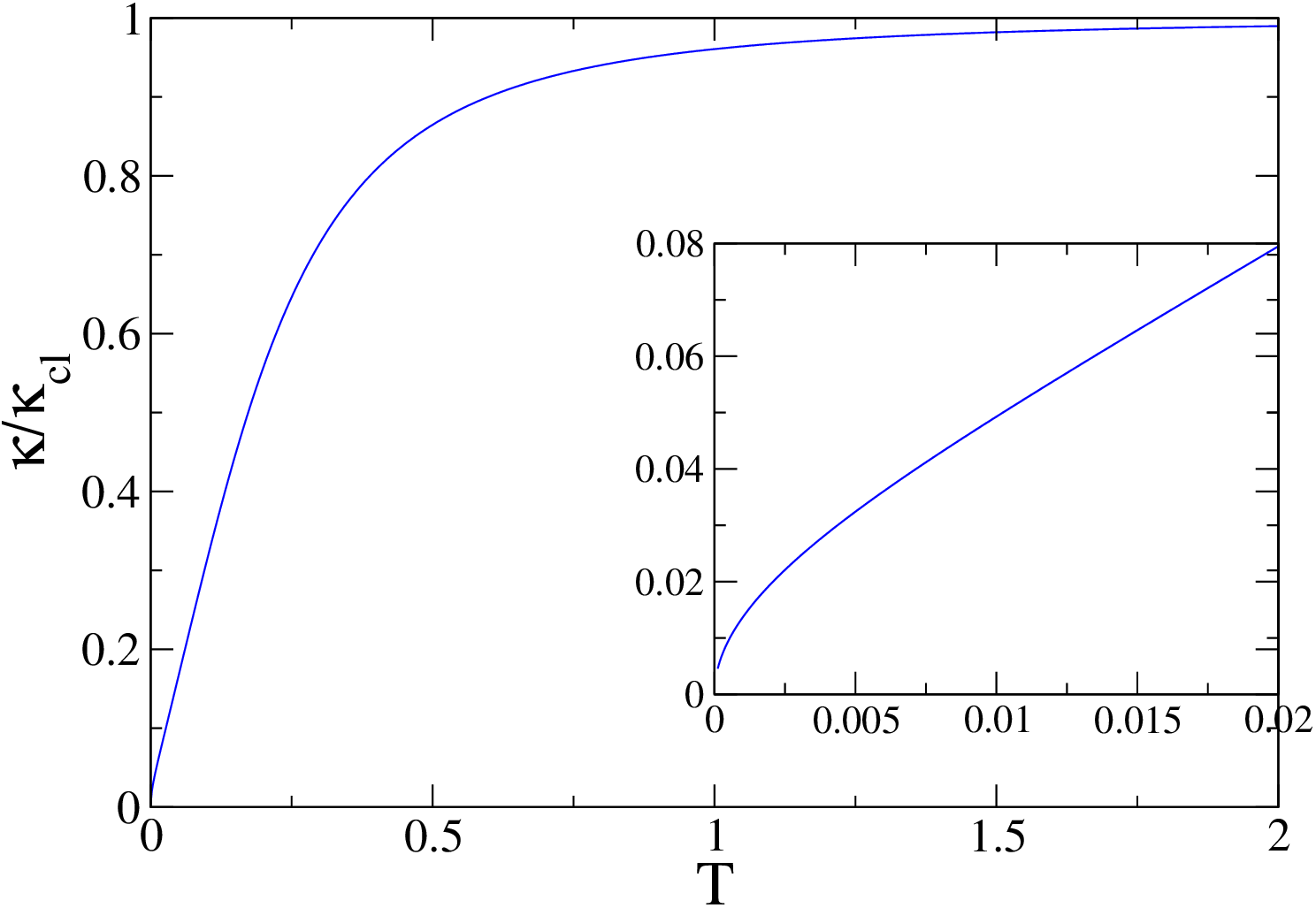}
\caption{ Plot of the scaled thermal conductivity as a function of
  temperature (in units of $\hbar \om_c/k_B$)  for $\nu=0.0$. Inset
  shows the low temperature behaviour. } 
\label{fig2}
\end{figure}

Finally we examine the general case where the coupling constant has a
finite value. 
In Figs.~(\ref{fig1},\ref{fig2}) we plot the thermal conductivity as a
function of temperature for two sets of parameter: (i)
$\g/(m\om_c)=0.2, \nu=0.5$  and (ii) $\g/(m \om_c) = 0.2,
\nu=0.0$. The insets in the two figures show the low-temperature
behaviour. As before, the low-temperature behaviour depends on whether or not
there is an onsite potential. However we see that the form of the
low-temperature behaviour is very different from the case of weak
coupling. For small 
$T$ it is easy to pull out the 
temperature dependence of the integral in Eq.~(\ref{genF}) and we find that
$\kappa \sim T^{3}$ for $\nu\neq 0$ and $\kappa \sim T^{1/2}$ for $\nu =
0$, in agreement with the numerical result shown in
Figs.~(\ref{fig1},\ref{fig2}).


\section{Discussion}
\label{sec:Disc}

In this paper, using the quantum Langevin equation approach,  we have
derived NEGF-like expressions for the heat current in a harmonic
lattice connected to external reservoirs at different temperatures. We
note that unlike other approaches such as the Green-Kubo formalism
and Boltzmann equation approach, the Langevin equation approach explicitly
includes the reservoirs. The Langevin equation is also physically
appealing since it gives a nice picture of the reservoirs as sources
of noise and dissipation. Also just as the Landauer formalism and NEGF
have been extremely useful in understanding electron transport in
mesoscopic systems it is likely that a similar description will be
useful for the case of heat transport in insulating nanotubes,
nanowires, etc. We note here that the single-channel Landauer results
follow from NEGF if one considers one-dimensional reservoirs
\cite{dhar03} and have been useful in interpreting experimental results
\cite{schwab00}.   

We think that the similarity, between heat conduction studies in 
harmonic systems and electron transport in noninteracting wires, is an
interesting and useful point to note. The two areas have developed quite
independently using different theoretical tools. In the former case
most of the earlier studies were done on classical systems using
either a Langevin or a Fokker Planck description. More recent studies
on quantum systems have used either a quantum Langevin or a density
matrix approach. On the other hand, for the electron case, which is
inherently quantum-mechanical, the most popular and useful approach
has been the Landauer and the NEGF formalism. As we have demonstrated,
in this paper for the phonon case, and in Ref~\cite{dhar06} for
the electron case, the NEGF results can be easily derived using the
Langevin approach at least in the noninteracting case. 

As an application of the NEGF results we have studied the problem of
heat conduction in  a one dimensional harmonic chain connected to
self-consistent reservoirs.  The classical version of this problem was
solved exactly by Bonetto et al \cite{bonetto04}, using 
different methods. The quantum mechanical case was studied earlier by
Vischer and Rich \cite{viss75b} and they obtained the thermal
conductivity in the limit of weak coupling. 
The advantage of the present approach is that it's implementation is
simple and straightforward. We obtain a general expression for the
thermal conductivity, which in limiting cases gives both the classical
result in \cite{bonetto04} and the weak-coupling result in \cite{viss75b}. 
We also find that, at low temperatures, the temperature dependence of
thermal conductivity in the case of finite coupling is completely
different from the weak coupling case.
In the classical case it was shown in \cite{bonetto04} that the
thermal conductivity of the harmonic chain with self-consistent
reservoirs can also be obtained from the Kubo formula. 
An interesting problem would be to demonstrate this in the
quantum-mechanical case.
It is
interesting to note that the self-consistent reservoirs are 
very similar to the Buttiker probes 
\cite{butt85,butt86} which have been used to model  inelastic
scatterering in electron transport. In the
electron case they lead to Ohm's 
law being satisfied just as in the harmonic  chain the introduction of 
self-consistent reservoirs leads to Fourier's law being satisfied.  
In fact we have recently shown how one can obtain Ohm's law using
self-consistent reservoirs modeled microscopically by noninteracting
electron baths \cite{roy06}. 

The Langevin equation approach has some advantages. 
For example, in the classical heat conduction case,  
it is easy to write Langevin equations for nonlinear systems and study
them numerically. Also they might be useful in studying time dependent
phenomena.  Examples of this are the treatment of quantum pumping in
\cite{hanggi05} and the treatment of the question of approach to the 
non-equilibrium  steady state in \cite{dhar06}. 
We feel that it is worthwhile to explore the possibility of using the
quantum Langevin equation approach to the harder and more interesting
problems  involving interactions and time-dependent potentials in both
the electron and phonon case.

\section*{Acknowledgments}

A.D thanks Diptiman Sen and Joseph Samuel for useful discussions. A.D
also thanks Keijo Saito for pointing out Refn.~\cite{viss75b}.

\appendix

\section{Green's function properties}
\label{app:GF}

We will consider some properties of the Phonon Green's functions. We denote by
$\CG^+(t)$ the full Green's function of the coupled system of wire and 
reservoirs. Let $U$ and $\Omega^2$ respectively denote the normal mode eigenvector
and eigenvalue matrices satisfying the equations:
\bea
U^T \Phi U =  \Omega^2,~~~~U^T M U= \hat{I} ~.
\eea 
We define the Green's function $\CG^+(t)$  as
\bea
\CG^+(t)= U~ \f{\sin{(\Omega t)}}{\Omega}~ U^{T}~ \theta(t) ~.
\eea
It satisfies the equation
\bea
M \ddot{\CG}^+(t)+ \Phi~ \CG^+(t)= \delta(t) ~\hat{I} ~.
\eea
The Fourier transform $\CG^+(\omega)=\int_{-\infty}^\infty dt ~\CG^+(t) e^{i \omega
t}$ is thus given by
\bea
\CG^+(\omega)= \f{1}{-(\omega+i \e)^2 M + \Phi} ~.
 \eea
The isolated reservoir Green's functions are given by:
\bea
g^+_L(\omega)&=&\f{1}{-(\omega+i \e)^2 M_L+ \Phi_L} \nn \\ 
g^+_R(\omega)&=&\f{1}{-(\omega+i \e)^2 M_R + \Phi_R}~. \nn
\eea
We can also represent $\CG^+(\omega)$ as follows:
\bea
&&\CG^+_{rs}(\omega)= -\sum_{Q} \f{U_{r Q}U_{s Q} }{(\omega+\omega_Q+i
\e)(\omega-\omega_Q+i\e)} \nn \\
&=& -\sum_Q \f{U_{r Q}U_{s Q} }{\omega^2-\omega_Q^2}+\f{i \pi}{2 \omega} \sum_Q
U_{r Q}U_{s Q}  [\d (\omega-\omega_Q) + \d (\omega + \omega_Q)] ~.
\eea
We will now express the
wire-part of the full Green's function in terms of the uncoupled
reservoir Green's functions. We  write the equation for $\CG^+(\omega)$
in the following form:
\bea
&&\left( \begin{array}{ccc} 
-M_W~ (\omega+i \e)^2 \hat{I}+ \Phi_W & V_L &  V_R \cr
 V_L^T& -M_L~ (\omega+i \e)^2 \hat{I}+ \Phi_L & 0 \cr
 V_R^T & 0 & -M_R ~(\omega+i \e)^2 \hat{I}+\Phi_R \cr
\end{array} \right) \nn \\
&& \times ~~ \left( \begin{array}{ccc} 
G^+_{W} & G^+_{WL} & G^+_{WR} \cr
G^+_{LW} & G^+_L & G^+_{LR} \cr
G^+_{RW} & G^+_{RL} & G^+_R \cr
\end{array} \right) ~
=~ \left( \begin{array}{ccc} 
\hat{I} & 0 & 0 \cr
0 & \hat{I} & 0 \cr
0 & 0 & \hat{I} \cr
\end{array} \right) ~.
\eea
{}From these equations we obtain the following expression for $G^+_W(\omega)$:
\bea
G^+_W(\omega)&=&\f{1}{- (\omega+i \e)^2~M_W + \Phi_W - 
\S_L^+ -  \S_R^+} ~, \label{green}\\
{\rm where} \quad \S^+_L(\omega) &=& V_L~ g^+_L(\omega)~  ~V_L^T ~, \nn \\ 
\S_R^+(\omega) &=&V_R~ g_R^+(\omega)~  V_R^T ~. \nn 
\eea

\section{Equilibrium properties}
\label{app:Equil}

In this section we will calculate the canonical ensemble expectation value of
$K=\la~\dot{X}_W \dot{X}_W^T~\ra$ where the average is taken over the 
equilibrium density matrix of the entire coupled system of wire and 
reservoirs. Denoting by $Z_Q$ the normal mode coordinates of the
entire system we get, for points $i,j$ on the wire: 
\bea
K^{eq}_{ij}&=&\la \dot{X}_i \dot{X}_j \ra_{eq} \nn \\
&=& \sum_Q U_{iQ} U_{jQ} \la \dot{Z}_Q^2 \ra_{eq} \nn \\
&=& \sum_Q U_{iQ} U_{jQ} [\f{\hbar \omega_Q}{2} + \hbar \omega_Q f(\omega_Q,T)] \nn \\
&=& \int_{-\infty}^\infty d \omega \f{\omega}{\pi} \sum_Q U_{iQ} U_{jQ} \f{\pi}{2\omega}
[\d (\omega-\omega_Q)+\d (\omega+\omega_Q)] \f{\hbar \omega}{2} \coth({\f{\hbar \omega}{2 k_B T}}) 
\nn \\
&=& \int_{-\infty}^\infty d \omega \f{\omega}{2\pi i } [(G^+_W-G^-_W) ]_{ij} 
\f{\hbar \omega}{2} \coth ({\f{\hbar \omega}{2 k_B T}}) ~.
\eea
Now from Eq.~(\ref{green}) we have:
\bea
(G^-_W)^{-1}-(G^+_W)^{-1}&=&( \S_L^+-\S_L^-) + ( \S_R^+ -
\S_R^-) +4 i \e \omega M_W \nn \\
&=&2~i~  (\G_L +\G_R) +4 i \e \omega M_W \nn \\
\Rightarrow G^+_W-G^-_W &=& 2~i~ G^+_W ~ (\G_L +\G_R)~ G^-_W +4 i \e 
\omega~ G_W^+ M_W G_W^- ~.~~~~~~~~
\eea
Hence we finally get:
\bea
K^{eq}_{ij}&=&\int_{-\infty}^\infty d \omega~ \f{\omega}{\pi }~ [~ G^+_W~  
(\G_L + \G_R)~ G^-_W~ ]_{ij}~ \f{\hbar \omega}{2}~ 
\coth ({\f{\hbar \omega}{2 k_B T}}) \nn \\
&& + \int_{-\infty}^\infty d \omega~ \f{2\e \omega}{\pi}~[~ G^+_W ~M_W~G^-_W~ 
]_{ij} ~\f{\hbar \omega}{2}~ \coth ({\f{\hbar \omega}{2 k_B T}}) ~.
\label{eqK}
\eea
Since we eventually take the limit $\e \to 0$, the second term is
non-vanishing only when the equation 
\bea
Det[-\omega^2 M_W+ \Phi_W- \S_L^+(\omega)-\S_R^+(\omega)] = 0
\eea
has solutions for real $\omega$. These solutions correspond to the
\emph{bound states}\cite{ziman72} of the coupled system of wire and reservoirs
\cite{dhar06}.


\begin{thebibliography}{10}
\bibitem{ziman72} J. M. Ziman, pp 71, {\it Principles of the theory of solids}
  (Second Edition, Cambridge University Press, 1972).

\bibitem{bonetto00}
F.~Bonetto, J.~L. Lebowitz, and L.~Rey-Bellet,
{\it {F}ourier's law: a challenge to theorists\/}.
\newblock In A.~Fokas, A.~Grigoryan, T.~Kibble, and B.~Zegarlinski
(eds.), {\it   Mathematical Physics 2000\/}, pp. 128--150, London,
2000. Imperial College  Press.

\bibitem{lepri01}
S.~Lepri, R.~Livi, and A.~Politi, {\it Thermal conduction in classical
  low-dimensional lattices\/}, Phys. Rep.~{\bf 377}, 1 (2003).

\bibitem{fourier}   T. Hatano, {\it Heat
  conduction in the diatomic    Toda lattice revisited},
  Phys. Rev. E {\bf 59}, R1 (1999);  
A. Dhar, {\it Heat conduction in
  a one-dimensional gas of elastically colliding particles of unequal
  masses}, Phys. Rev. Lett. {\bf 86}, 3554 (2001); B. Li, H. Zhao and
  B. Hu, {\it Can Disorder Induce a Finite Thermal    Conductivity in
  1D Lattices? }, Phys. Rev. Lett. {\bf 86}, 63 (2001);  
 P. Grassberger, W. Nadler and L. Yang, {\it Heat Conduction and
  Entropy Production in a One-Dimensional Hard-Particle Gas},
Phys. Rev. Lett. {\bf 89}, 180601 (2002); A. V. Savin, G. P. Tsironis
  and A. V. Zolotaryuk, {\it Heat conduction 
  in one-dimensional systems with   hard-point interparticle
  interactions}, Phys. Rev. Lett. {\bf 88}, 154301 (2002);  
 G. Casati and T. Prosen, {\it Anomalous heat conduction in a one-dimensional
  ideal gas},  Phys. Rev. E {\bf 67}, 015203(R) (2003); 
J. S. Wang and B. Li,{\it   Intriguing Heat Conduction of a Chain with
  Transverse Motions}, Phys. Rev. Lett. {\bf 92}, 074302 (2004).


\bibitem{lepri97} S. Lepri, R. Livi, and A. Politi, {\it Heat
  Conduction in Chains of Nonlinear   Oscillators},
  Phys. Rev. Lett. {\bf78}, 1896 (1997); 


\bibitem{narayan02} O. Narayan and S. Ramaswamy, {\it Anomalous heat
  conduction in   one-dimensional momentum-conserving systems},
  Phys. Rev. Lett. {\bf   89}, 200601 (2002). 


\bibitem{dhar01} A. Dhar, {\it Heat conduction in the disordered
  harmonic chain revisited}, Phys. Rev. Lett. {\bf 86}, 5882 (2001).

\bib{datta} S. Datta, {\it Electronic transport in mesoscopic systems} 
(Cambridge University Press, 1995).

\bib{imry} Y. Imry, {\it Introduction to Mesoscopic Physics} (Oxford
University Press, 1997).

\bib{spohn05} H. Spohn, {\it The phonon Boltzmann equation, properties
  and link to weakly anharmonic lattice dynamics}, math-ph/0505025/.

\bibitem{rieder67}
Z.~Rieder, J.~L. Lebowitz, and E.~Lieb,
{\it Properties of a harmonic crystal in a stationary nonequilibrium state\/},
J. Math. Phys.~{\bf 8}, 1073 (1967).

\bibitem{rubin71} R. Rubin and W. Greer, {\it Abnormal Lattice Thermal
  Conductivity of a One-Dimensional, Harmonic, Isotopically Disordered
  Crystal}, J. Math. Phys. (N.Y.) {\bf 12},  1686 (1971).


\bibitem{connor74} A. J. O'Connor and J. L. Lebowitz, {\it Heat
  conduction and sound transmission in isotopically disordered
  harmonic crystals}, J. Math. Phys.{\bf 15},  692 (1974).


\bibitem{lee05} L. W. Lee and A. Dhar, {\it Heat Conduction in a
  two-dimensional harmonic crystal with   disorder},
  Phys. Rev. Lett. {\bf 95}, 094302 (2005).

\bibitem{hu87} G. Y. Hu and R. F. O'Connell, {\it Quantum transport
  for a many-body system using a quantum Langevin-equation approach},
  Phys. Rev. B {\bf 36}, 5798 (1987). 

\bibitem{clel92} A. N. Cleland, J. M. Schmidt and J. Clarke, {\it
  Influence of the environment on the Coulomb blockade in
  submicrometer normal-metal tunnel junctions},  Phys. Rev. B {\bf
  45}, 2950 (1992); G. Y. Hu and R. F. O'Connell, {\it  Charge
  fluctuations and zero-bias resistance in small-capacitance   tunnel
  junctions},  Phys. Rev. B {\bf 49}, 16505 (1994), Phys. Rev. B 46,
  14219 (1992). 


\bibitem{chen89} 
    Y.-C. Chen, J. L. Lebowitz, and C. Liverani, {\it Dissipative
    quantum dynamics in a boson bath},   Phys. Rev. B {\bf 40},
    4664 (1989).

\bib{zurcher90} U. Zurcher and P. Talkner, {\it Quantum-mechanical
  harmonic chain attached to heat baths. II. Nonequilibrium
  properties}, Phys. Rev. A {\bf 42}, 3278 (1990).

\bibitem{saito00} K. Saito, S. Takesue, and S. Miyashita, {\it Energy
 transport in the integrable system in contact with various types of
 phonon reservoirs},  Phys. Rev. E {\bf 61}, 2397 (2000).

\bibitem{dhar03} A. Dhar and B. S. Shastry, {\it Quantum transport
  using the Ford-Kac-Mazur formalism}, Phys. Rev. B {\bf 67}, 195405 (2003).

\bibitem{segal03} D. Segal, A. Nitzan and P. Hanggi, {\it Thermal
  conductance through molecular wires}, J. Chem. Phys. {\bf
  119},  6840  (2003).

\bibitem{dhar06} A. Dhar and D. Sen, {\it Nonequilibrium Green's
  function formalism and the problem of bound states},   Phys. Rev. B
  {\bf 73}, 085119 (2006). 

\bibitem{sheng} P. Sheng, {\it Introduction to wave scattering,
  localization and mesoscopic phenomena} (Academic Press, 1995).

\bibitem{feng} R. Berkovits and S. Feng, {\it Correlations in coherent
  multiple scattering}, Phys. Reps. {\bf 238}, 135 (1994).


\bibitem{rego98} L. G. C. Rego and G. Kirczenow, {\it Quantized
  Thermal Conductance of Dielectric Quantum Wires},
  Phys. Rev. Lett. {\bf 81}, 232 (1998). 

\bibitem{blencowe99}M. P. Blencowe, {\it Quantum energy flow in
  mesoscopic dielectric   structures},   Phys. Rev. B {\bf 59},
  4992 (1999). 

\bibitem{yamamoto06} T. Yamamoto and K. Watanabe, {\it Nonequilibrium
  Green's function approach to phonon transport in defective carbon
  nanotubes}, Phys. Rev. Lett. {\bf 96}, 255503 (2006).

\bib{meir92} Y. Meir and N. S. Wingreen, {\it Landauer formula for the
  current through an interacting electron region},
  Phys. Rev. Lett. {\bf 68}, 2512 (1992).  

\bibitem{kamenev04} A. Kamenev, in {\it Nanophysics: Coherence and Transport}
(Lecture notes of the Les Houches Summer School 2004).


\bibitem{viss70}
M.~Bolsterli, M.~Rich, and W.~M. Visscher,
{\it Simulation of nonharmonic interactions in a crystal by self-consistent
  reservoirs\/}, Phys. Rev. A {\bf  4}, 1086 (1970).

\bibitem{viss75}
M.~Rich and W.~M. Visscher,
{\it Disordered harmonic chain with self-consistent reservoirs\/},
Phys. Rev. B {\bf 11}, 2164 (1975).

\bibitem{viss75b}
W.~M. Visscher and M. Rich, {\it Stationary nonequilibrium properties
  of a quantum-mechanical lattice with self-consistent reservoirs},
Phys. Rev. A {\bf 12}, 675 (1975). 

\bibitem{bonetto04} F. Bonetto, J. L. Lebowitz, and J. Lukkarinen,
  {\it Fourier's law for a harmonic crystal with self-consistent
  reservoirs}, J. Stat. Phys. {\bf116}, 783 (2004).

\bibitem{schwab00} K. Schwab, E. A. Henriksen, J. M. Worlock and
  M. L. Roukes, {\it Measurement of the quantum of thermal
  conductance}, Nature {\bf 404}, 974 (2000).  

\bibitem{butt85} M. Buttiker, {\it Small normal-metal loop coupled to an
  electron reservoir}, Phys. Rev. B {\bf 32}, 1846 (1985).

\bibitem{butt86} M. Buttiker, {\it Role of quantum coherence in series
  resistors}, Phys. Rev. B {\bf 33}, 3020 (1986). 

\bibitem{roy06} D. Roy and A. Dhar, {\it  Electron transport in a one dimensional conductor with inelastic scattering by self-consistent reservoirs},
 Phys. Rev. B {\bf 75}, 195110 (2007). 

\bibitem{hanggi05} M.Strass, P. Hanggi and S. Kohler,
{\it Nonadiabatic electron pumping: maximal current with minimal noise},
 Phys. Rev. Lett. {\bf 95}, 130601 (2005). 
\end{thebibliography}
\end{document}